\documentclass[12pt]{article}

\def\be{\begin{equation}}
\def\ee{\end{equation}}
\def\bea{\begin{eqnarray}}
\def\eea{\end{eqnarray}}
\def\bean{\begin{eqnarray*}}
\def\eean{\end{eqnarray*}}

\newtheorem{theorem}{Theorem}

\begin{document}

\title{A local potential for the Weyl tensor in all dimensions}
\author{S. Brian Edgar$^1$ and Jos\'e M.M. Senovilla$^2$
\\ $^1$ Department of Mathematics,ææ Link\"{o}pings universitet\\ 
Link\"{o}ping, Sweden S-581 83\\
$^2$ F\'{\i}sica Te\'orica, Universidad del Pa\'{\i}s Vasco, \\
Apartado 644, 48080 Bilbao, Spain\\ 
e-mails: bredg@mai.liu.se, wtpmasej@lg.ehu.es}

\maketitle

\begin{abstract} 
In all dimensions $n\ge 4$ and arbitrary signature, we demonstrate the
existence of a new local potential --- a double $(2,3)$-form,
$P^{ab}{}_{cde}$ --- for the Weyl curvature tensor $C_{abcd}$, and
more generally for all tensors $W_{abcd}$ with the symmetry properties
of the Weyl tensor.  The classical four-dimensional Lanczos potential
for a Weyl tensor --- a double $(2,1)$-form, $H^{ab}{}_{c}$ --- is
proven to be a particular case of the new potential: its double dual.
\end{abstract}

\

\


\


\

\

In $n$ dimensions, the existence of a $1$-form potential $A_{a}$ for 
the $2$-form electromagnetic field $F_{ab}$ enables the 
electromagnetic field equations to be written as a wave equation (in 
Lorentz signature) for 
the potential, which is particularly simple in the differential 
gauge\footnote{A semicolon indicates covariant derivative with respect 
to the canonical connection. As usual, we use round and square brackets 
to denote symmetrization and antisymmetrization of indices, respectively.
Our convention for the Riemann tensor follows from the 
Ricci identity: $2v_{a;[bc]}=R^d{}_{abc}v_{d}$.} 
$A^a{}_{;a}=0$ \cite{MTW}:
$$
\nabla^2A^a=F^{ab}{}_{;b} \, .
$$
In {\it four} dimensions, Lanczos \cite{L} proposed the existence of a
double $(2,1)$-formææ potential $H^{ab}{}_{c}=H^{[ab]}{}_{c}$ for the
double $(2,2)$-form Weyl tensor $C^{ab}{}_{cd}$ (see e.g. \cite{S} for
the definition and properties of $r$-fold forms), and this result was
confirmed in \cite{BC} (for {\it any} double (2,2)-form with the
algebraic properties of the Weyl conformal curvature tensor) and
equivalently for {\it any} symmetric spinor $\phi_{ABCD}$ in \cite{I};
see also \cite{AE1}, \cite{EH1}.ææ

It is straightforward to conjecture a direct $n$-dimensional analogue
for the Lanczos-Weyl equation (see (\ref{WnH}) below) but,
unfortunately, it has been shown that such a potential cannot exist,
in general, in dimensions $n>4$ \cite{EH2}.  As a consequence,
interest in the existence of potentials for the Weyl tensor in
dimensions $n>4$ has diminished.

However there are a number of reasons for a continuing interest in a
potential for the Weyl tensor, especially one which is defined in
arbitrary dimensions.  In particular, one attractive property of any
possible potential for the Weyl tensor is that it has units $L^{-1}$, 
the same as the connection (or the first derivative of the metric). 
This means that any of its squares (such as its superenergy tensor \cite{S})
would have units $L^{-2}$ which are precisely the units we would expect 
for quantities related to gravitational `energies', in contrast with
the familiar Bel-Robinson superenergy tensor \cite{BR,S} whose units
are $L^{-4}$ 
(Roberts \cite{R1} pointed this out for the Lanczos potential).
As a matter of fact, another related attractive property of these
potentials is that they are at a level similar to the connection. As 
the potential is an `integral' of the Weyl tensor, which itself is a second 
derivative of the metric, the potential and the connection are 
necessarily connected. The advantage is that the potential is a 
tensorial object. Then, of course, the impossibility of constructing 
tensors from the metric and its first derivatives manifest itself 
---as it must--- in the lack of uniqueness of the potentials. 
Thus, already at this stage we realize that any potential must be affected by 
a {\em gauge freedom}.

\

In this letter we shall show, in arbitrary dimension and signature,
that all $(2,2)$-forms with the algebraic properties of the Weyl
tensor, locally have a double (2,3)-form potential
$P^{ab}{}_{cde}=P^{[ab]}{}_{[cde]}$.  We will further show that, in
four dimensions, the double Hodge dual of the new potential, which is
a double (2,1)-form, is {\em exactly} the classical Lanczos potential. 
Hence a very natural and very general potential for the Weyl tensor is
this new double $(2,3)$-form potential.

\

We begin by considering an arbitrary $n$-dimensional pseudo-Riemannian
manifold with metric $g_{ab}$ of any signature.  We call a {\em Weyl
candidate} any double (2,2)-form $X_{abcd}=X_{[ab]cd}=X_{ab[cd]}$ with
the algebraic properties of the Weyl conformal curvature tensor:
\be
X^a{}_{bca}=0, \hspace{5mm} X_{a[bcd]}=0 \,\,\,
(\Longrightarrow \,\,\, X_{abcd}=X_{cdab}),\label{v-prop}
\ee
so that $X_{abcd}$ is a {\em traceless and symmetric} double (2,2)-form. 
We will now exploit these properties and rearrange as 
follows,
$$
\nabla^2X_{abcd}= \frac{1}{2}\left(X_{abcd;e}{}^e+X_{cdab;e}{}^e\right)
=\frac{1}{2}\left(3X_{ab[cd;e]}{}^e-2 X_{abe[c;d]}{}^{e} 
+3X_{cd[ab;e]}{}^e-2 X_{cde[a;b]}{}^{e}\right) 
$$
and using the Ricci identity on the second and fourth terms we obtain
\bea
\nabla^2 X_{abcd}= \frac{1}{2}\left( 3X_{ab[cd;e]}{}^e-2 X_{abe[c}{}^{;e}{}_{d]} 
+3X_{cd[ab;e]}{}^e-2 X_{cde[a}{}^{;e}{}_{b]}\right)+\hspace{15mm} \nonumber\\ 
2\left(R^e_{dn[a}X^n{}_{b]ec}-R^e_{cn[a}X^n{}_{b]ed}\right)+
R_{n[c}X^n{}_{d]ab}+R_{n[a}X^n{}_{b]cd}-
\frac{1}{2}\left(R_{cden}X_{ab}{}^{en}+R_{aben}X_{cd}{}^{en}\right) \label{rxv}
\eea
where $R_{ab}$ is the Ricci tensor. Using the shorthand notation
$\{R \otimes X\}_{abcd}$ for all the terms linear in the curvature 
tensors, that is to say, the second line of (\ref{rxv}), we can 
rewrite by using  (\ref{v-prop}) repeatedly
\begin{eqnarray}
\nabla^2 X_{abcd}&=&\frac{1}{2}\left(3X_{ab[cd;e]}{}^e+
4 X_{e[ab][c}{}^{;e}{}_{d]} 
+3X_{cd[ab;e]}{}^e+4 X_{e[cd][a}{}^{;e}{}_{b]}\right)+\{R \otimes X \}_{abcd} 
\nonumber\\
&=&\frac{3}{2}\left(X_{ab[cd;e]}{}^e+X^e{}_{a[bc}{}_{;e]}{}_{d}- 
X^e{}_{b[ac}{}_{;e]}{}_{d} + X^e{}_{b[ad}{}_{;e]}{}_{c} - 
X^e{}_{a[bd}{}_{;e]}{}_{c}+ X_{cd[ab;e]}{}^e\right.  \nonumber\\æææ
&& + \left.  X^e{}_{c[da}{}_{;e]}{}_{b}- 
X^e{}_{d[ca}{}_{;e]}{}_{b}+ X^e{}_{d[cb}{}_{;e]}{}_{a}-
X^e{}_{c[db}{}_{;e]}{}_{a }\right)+\{R \otimes X\}_{abcd}. \label{X} 
\end{eqnarray}

Now define the double $(2,3)$-form 
\be
P_{abcde}=P_{[ab][cde]}\equiv {3\over 2} X_{ab[cd;e]}\label{P=X}
\ee
which will inherit from $X_{abcd}$ the properties
\be
P_{a[bcde]}=0, \hspace{1cm} P^{ab}{}_{abc}=0 . \label{Ppty1}
\ee
Immediate consequences from the first of these are the following 
useful properties 
\be
P^e{}_{[bcd]e}=0, \hspace{5mm} P_{[abcd]e}=0, \hspace{5mm}
P_{abcde}=3 P_{[cde]ab}=3 P_{a[cde]b}, 
\hspace{5mm} P_{a[bc]de}=-P_{a[de]bc}\label{Ppty2} \, .
\ee
Hence (\ref{X}) can be rewritten as
\begin{eqnarray}
\nabla^2 X_{abcd} &=& P_{abcde}{}^{;e}+P_{cdabe}{}^{;e}+ 
4P_{e[ab][c}{}^{e}{}_{;d]}+ 
4P_{e[cd][a}{}^{e}{}_{;b]} + \{R \otimes X \}_{abcd} \nonumber\\ 
&=& P_{abcde}{}^{;e}+P_{cdabe}{}^{;e}-2 
P^e{}_{[c|abe|;d]}-2P^e{}_{[a|cde|;b]}+ \{R \otimes X\}_{abcd} \, .
\end{eqnarray}
It is easily confirmed that this expression constructed from 
$P_{abcde}$, ignoring the $\{R \otimes X\}_{abcd}$ terms,
has the necessary Weyl candidate index symmetries.

Now consider {\it any} Weyl candidate $W_{abcd}$.
We can always find a Weyl candidate `superpotential' $X_{abcd}$ 
locally for $W_{ab}{}_{cd}$ by 
appealing to the Cauchy-Kowalewsky theorem \cite{CH} which guarantees a local 
solution of the linear second order equation
\be
\nabla^2X_{abcd}- \{R \otimes X \}_{abcd} = W_{abcd} \label{boxX=W}
\ee
in a given analytic background space. From the superpotential $X_{abcd}$ 
we can then construct the potential $P_{abcde}$ using (\ref{P=X}), 
and obtain our main result:
\begin{theorem}
Any Weyl candidate tensor field
$W_{abcd}$ has a double $(2,3)$-form local potential $P_{abcde}$ 
with the properties (\ref{Ppty1}) such that
\be 
W^{ab}{}_{cd} = P^{ab}{}_{cdi}{}^{;i}+P_{cd}{}^{abi}{}_{;i} -2 
P_i{}_{[c}{}^{abi}{}_{;d]}-2P^i{}^{[a}{}_{cdi}{}^{;b]} \, . \label{W=P}
\ee
The potential itself can be given in terms of a Weyl candidate
superpotential $X_{abcd}$ by (\ref{P=X}) so that $W_{abcd}$ is given 
in terms of this superpotential by (\ref{boxX=W}).
\end{theorem}
Of course the  above theorem has a direct application to the 
Weyl tensor $C^{ab}{}_{cd}$ of any pseudo-Riemannian manifold.

The number of independent components of the potential can be 
computed easily by using the properties (\ref{Ppty1}) and the result is
$(n+2)n(n-3)(n^2-n+4)/24$ (16 if $n=4$, 70 if $n=5$). This is 
larger than the number of independent components of a Weyl 
candidate, which is known to be $(n+2)(n+1)n(n-3)/12$ (that is, 10
if $n=4$, 35 if $n=5$). It is also larger (equal, in the case $n=4$) than the number of 
independent Ricci rotation coefficients, or of independent components 
of the connection in a given basis. Although this large number of 
components may seem unsatisfactory, one must bear in mind that this 
number can be substantially reduced in any particular case by means of 
the gauge differential freedom  as we show in \cite{ES}.

The choice of $P_{abcde}$ is not unique. As  is 
usual with potentials --- see, e.g., \cite{MTW} --- one can redefine 
significant parts of $P_{abcde;f}$ without altering the combination 
in (\ref{W=P}) which produces a particular $W_{abcd}$.
From known results about $p$-forms, and standard arguments on the
independence of the exterior differential versus the divergence, we
can easily identify and exploit some of the gauge freedom for the new
potential.  A detailed discussion on gauge, with explicit formulas for
the gauge redefinition of $P_{abcde}$, will be given in \cite{ES}.

\

Consider now the special case of $n=4$.  It has already been noted
that if $n=4$ the Weyl tensor, and more generally all Weyl candidates,
have a so-called Lanzcos potential \cite{L}, \cite{BC}, \cite{I},
\cite{AE1}.  This is a double (2,1)-form $H^{ab}{}_{c}=H^{[ab]}{}_{c}$
with the extra properties
\be
H_{[abc]}=0, \hspace{1cm} H_{ab}{}^b=0 \label{Hpty}
\ee
and all Weyl candidates have this type of potentials such that
\be 
W^{ab}{}_{cd}ææ= 2 H^{ab}{}_{[c;d]}+2H_{cd}{}^{[a;b]}ææ-2 
\delta^{[a}_{[c}\left(H^{b]}{}^e{}_{d];e}+H_{d]e}{}^{b];e}\right) .
\label{W=H}
\ee
What is the relation, if any, between the new potential $P_{abcde}$ 
and the Lanczos potential?

To answer this question, we first of all remark that a double 
(2,3)-form is equivalent, via dualization with the Hodge * operator, 
to a double (2,1)-form in $n=4$. For the formulas and conventions about the 
Hodge dual operator we refer the reader to \cite{S}---allowing for an extra 
sign depending on the signature of the space. In particular, 
for any {\em traceless} double (2,2)-form, and denoting the canonical volume 
element 4-form by $\eta_{abcd}=\eta_{[abcd]}$, we have \cite{S}
$$
(*W*)_{abcd}\equiv \frac{1}{4}\eta_{abef}\eta_{cdgh}W^{efgh} 
\hspace{5mm} \Longrightarrow \hspace{5mm} (*W*)_{abcd}=\epsilon \,
W_{abcd}
$$
where $\epsilon =\pm 1= $ sign(det($g_{ab}$)) is a sign depending on
the signature ($\epsilon =1$ in positive-definite metrics, and
$\epsilon =-1$ in the Lorentzian case).  Similarly, for any double
(2,3)-form $P_{abcde}$ we can write \cite{S}
$$
(*P*)_{abc}\equiv \frac{1}{12}\eta_{abef}\eta_{dghc}P^{efdgh}\hspace{4mm}
\Longrightarrow \hspace{4mm} P^{ab}{}_{cde}=6\epsilon\, 
(*P*)^{[a}{}_{[cd}\delta^{b]}_{e]} , \hspace{5mm} 
P^i{}_{cabi}=\epsilon\,(*P*)_{abc} .
$$
Observe that the properties (\ref{Ppty1}) translate for the double 
dual into, respectively, 
$$
(*P*)_{[abc]}=0, \hspace{1cm} (*P*)_{ab}{}^b=0
$$
which are the Lanczos potential properties (\ref{Hpty}) exactly. 
Hence, by taking the double dual of (\ref{W=P}), using the previous 
formulas and after a little bit of algebra we can prove that, in four 
dimensions
\be
\epsilon\,ææW^{ab}{}_{cd}ææ= 2 (*P*)^{ab}{}_{[c;d]}+2(*P*)_{cd}{}^{[a;b]}ææ-2 
\delta^{[a}_{[c}\left((*P*)^{b]}{}^e{}_{d];e}+(*P*)_{d]e}{}^{b];e}\right) 
\label{W=PP}
\ee
which coincides with (\ref{W=H}) by identifying
$$
H_{abc}=\epsilon (*P*)_{abc} \, .
$$
 For completeness, we give also the inverse formula:
$P_{abcde}=\epsilon\, (*H*)_{abcde}$.

Therefore, we have recovered the classical Lanczos potential for the
Weyl tensor in four dimensions as the double dual of the new potential
$P_{abcde}$.  Note that the familiar differential gauge of the Lanczos
potential $H_{abc}{}^{;c}$, \cite{L}, \cite{BC}, \cite{I}, \cite{AE1}
becomesæ æ$H_{abc}{}^{;c}= \epsilon\, (*P*)_{abc}{}^{;c}$, and so the
differential gauge for the new potential resides in
$P^{ab}{}_{[cde;f]}$, via double dualization.

In dimensions greater than four, a natural generalization of
(\ref{W=H}) to arbitrary dimension is to keep the double (2,1)-form
$H_{abc}$ with properties (\ref{Hpty}) and consider the appropriate
formula analogous to (\ref{W=H}) which keeps the Weyl candidate
symmetries.  This formula is unique and reads
\be
W^{ab}{}_{cd}=2 H^{ab}{}_{[c;d]}+2H_{cd}{}^{[a;b]}-\frac{4}{n-2} 
\delta^{[a}_{[c}\left(H^{b]}{}^e{}_{d];e}+H_{d]e}{}^{b];e}\right)  \label{WnH} .
\ee
However, as already noted, this Lanczos potential exists exclusively
in $n=4$ dimensions \cite{EH2}.  On the other hand, we now know that
there is a different counterpart to the Lanczos potential in $n>4$
dimensions; it is the double dual of $P^{ab}{}_{cde}$, i.e., a double
$(2,n-3)$-form $H^{ab}{}_{c_1c_2\ldots c_{n-3}}$, and an expression
--- giving any Weyl candidate in terms of such a potential --- can be
obtained in any dimension by the same method that led to (\ref{W=PP}). 
Clearly such a dimensionally dependent result is much less natural and
convenient that our defining formula (\ref{W=P}), which is independent
of dimension.  Therefore, in particular, we have proved that {\em the
natural and proper way to consider a potential for a double (2,2)-form
is as a double (2,3)-form; this version carries over to all dimensions
by means of (\ref{W=P}).  The traditional double (2,1)-form Lanczos
potential $H^{ab}{}_c$ in $n=4$ dimensions is nothing but the double
dual of the general potential $P^{ab}{}_{cde}$}.

\

In electromagnetic theory the local existence of the potential $A_{a}$
for the field tensor $F_{ab}$ is a direct consequence of applying
Poincare's Lemma to one of Maxwell's equations, $F_{[ab;c]}=0$. 
However, it is important to note that the local existence of the new
potential $P_{abcde}$ for the Weyl tensor does not require any such
`field equations'.  Our result is actually analogous to the result
that any 2-form {\em always} has a pair of local potentials, a 1-form
$A_{a}$ and a 3-form $B_{abc}$ such that
$$
F_{ab}=2A_{[a;b]}-B_{abc}{}^{;c} \, .  
$$
If the 2-form $F_{ab}$ is closed, then $B_{abc}$ can be chosen to be
zero, while if it is divergence- free, then $A_{a}$ can be set to
zero, \cite{MTW}, \cite{I}.  Our result for Weyl candidates is the
generalization of this fact but, given the special structure of {\em
traceless and symmetric} double (2,2)-forms, we achieve dealing with
just {\em one} potential.  A detailed discussion will be presented in
\cite{ES}.

Finally we emphasise that all our results are local and depend on the
analiticity of the pseudo-Riemannian metric, \cite{CH}.  However, as
is usually the case, we expect that these results can be generalized,
by using appropriate techniques of existence and uniqueness of
solutions to differential equations, to the smooth case and even to
spaces of low differentiability; from the point of view of general
relativity, we can appeal to stronger theorems \cite{F}, \cite{HE}
when we specialise to spaces with Lorentz signature, and the second
order differentialææ equations in the theorems become wave
equations.ææ

\section*{Acknowledgements} 
JMMS gratefully acknowledges financial support from the Wenner-Gren Foundations, 
Sweden, and from grants BFM2000-0018 of the Spanish CICyT and 
no. 9/UPV 00172.310-14456/2002 of the University of the Basque 
Country. JMMS thanks the Applied Mathematics Department 
at Link\"oping University, where this work was carried out, 
for hospitality.

\end{document}